\documentclass{kapproc} 

\usepackage{procps} 

\usepackage[dvips]{graphicx}

\upperandlowercase

\kluwerbib 

\begin{document}

\articletitle{Emission line galaxies in clusters}


\author{Bianca M. Poggianti}
\affil{INAF-Osservatorio Astronomico di Padova - Italy\\
}
\email{poggianti@pd.astro.it}


\begin{abstract}
At the present epoch, clusters of galaxies are known to be a hostile
environment for emission-line galaxies, which are more commonly found
in low density regions outside of clusters. In contrast, going to
higher redshifts the population of emission-line galaxies in clusters
becomes progressively more conspicuous, and large numbers of
star-forming late-type galaxies are observed. I present an overview of
the observational findings and the theoretical expectations
regarding the evolution of emission-line galaxies in dense
environments, discussing the properties of these galaxies 
and the current evidence
for environmental influences on their evolution.
\end{abstract}


\section*{Introduction}

This review focuses on galaxies with emission lines in their optical
spectra. In the great majority of cases, at least in clusters, an
emission-line galaxy can be assumed to be currently forming
stars\footnote{In the following, any contribution to the emission
originating from an eventual AGN will be disregarded.}, thus the
subject of this contribution is essentially the whole population of
cluster galaxies with ongoing star formation at any given epoch. One
of the major goals of current studies of galaxy formation and
evolution is to understand when each galaxy formed or will form its
stars, and why.  Naturally, star-forming galaxies are at the heart of
this investigative effort. Knowing their frequency and their evolution
as a function of the environment is fundamental to trace the evolution
of the star formation activity with redshift and to start gaining
insight on the processes that regulate such activity.

This is an area of research that really encompasses ``many scales'' of
the Universe, from the star formation on the scale of a single
star-forming region within a galaxy, to the integrated properties on a
galaxy scale, to a galaxy-cluster scale and to the evolution of the
global star formation history on a cosmic scale.  I will briefly touch
upon each of these scales and I will focus on the following main
questions: how many emission line galaxies are there in clusters, at
different redshifts?  What are their properties (star formation
rates, Hubble types, gas content)? What is their fate?  How do they
evolve and why? And, finally, what can they teach us about galaxy
formation and evolution in general?  

\section{High redshift}

\subsection{Star formation activity}

At any given redshift, the properties of cluster galaxies
display a large cluster to cluster variance. The ``average
galaxy'' in the Coma cluster looks quite different from the ``average
galaxy'' in the Virgo cluster, the first cluster being dominated by
passively evolving, early-type galaxies, and the second one having a
larger population of star-forming spirals.  Moreover, clusters of
galaxies, far from being closed boxes, continuously ``form'' and
evolve accreting single galaxies, pairs, groups, or merging with other
clusters. The galaxy content of a cluster thus changes with redshift.
Keeping these two things in mind, it is possible to look for
evolutionary trends with redshift.

Historically, the first evidence for a higher incidence of
star-forming galaxies in distant clusters compared to nearby clusters
came from photometric studies (Butcher \& Oemler 1978, 1984, Ellingson
et al. 2001, Kodama \& Bower 2001).  Spectroscopy is of course the
most direct way to identify emission--line galaxies. For distant
galaxies, the $\rm H\alpha$ line is redshifted at optical wavelengths
that are severely affected by sky or in the near-IR, thus the feature
most commonly used is the [O{\sc ii}]$\lambda$3727 line.

In the MORPHS sample of 10 clusters at $z \sim
0.4-0.5$, the fraction of emission--line galaxies is $\sim 30$\% for
galaxies brighter than $M_V = -19 + 5 \rm log \, h^{-1}$ (Dressler et
al. 1999, Poggianti et al. 1999). In the CNOC cluster sample, at an
average redshift $z \sim 0.3$, this fraction is about 25\% (Balogh et
al. 1999).  This incidence is much higher than it is observed in
similarly rich clusters at $z=0$ (Dressler, Thompson \& Shectman 1988).
Significant numbers of emission-line galaxies have been reported in
virtually all spectroscopic surveys of distant clusters
(Couch \& Sharples 1987, Fisher et al. 1998, Postman et al. 1998, 2001).

\begin{figure}[ht]
\centerline{\includegraphics[width=8cm]{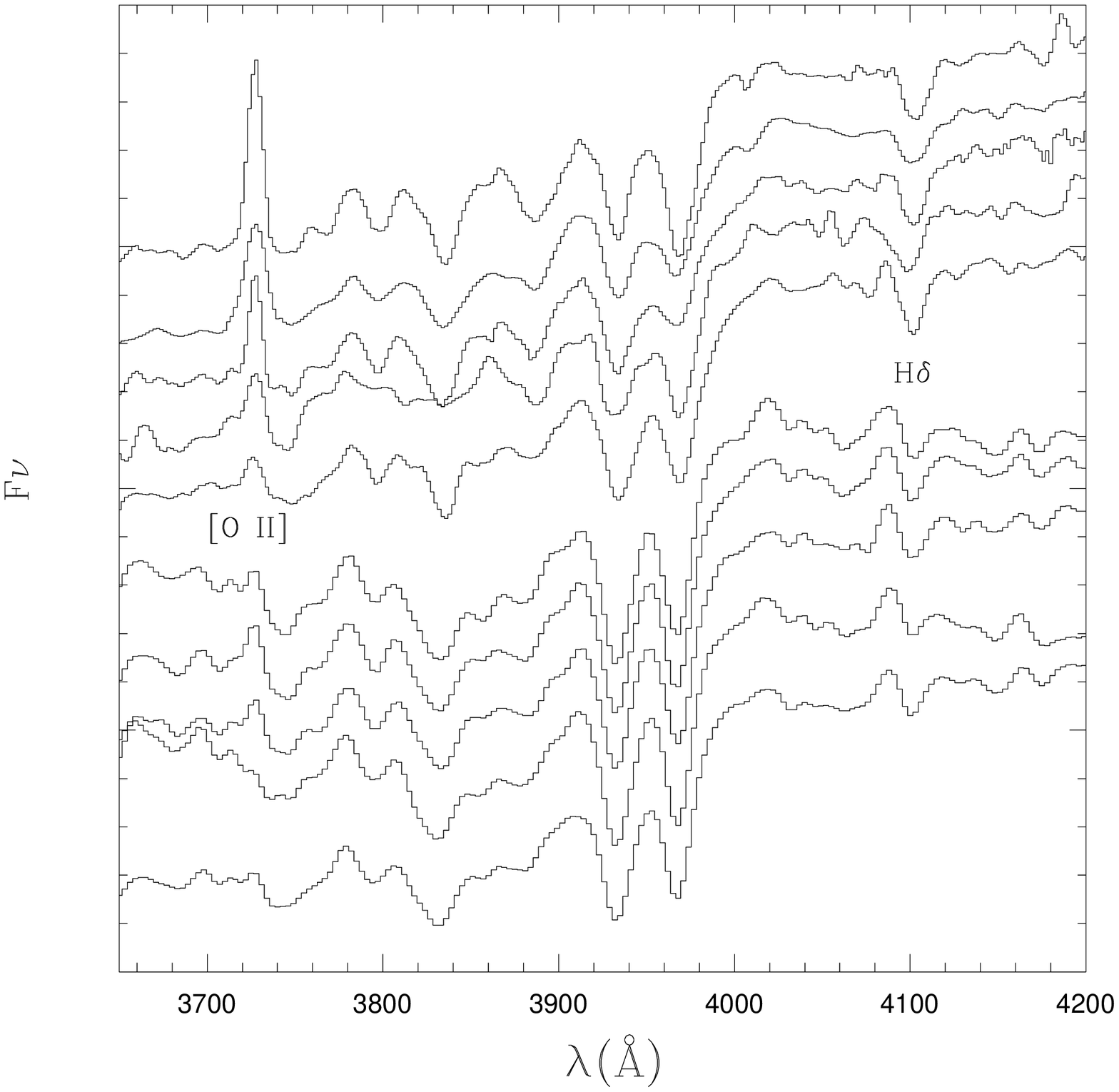}\hfill\includegraphics[width=8cm]{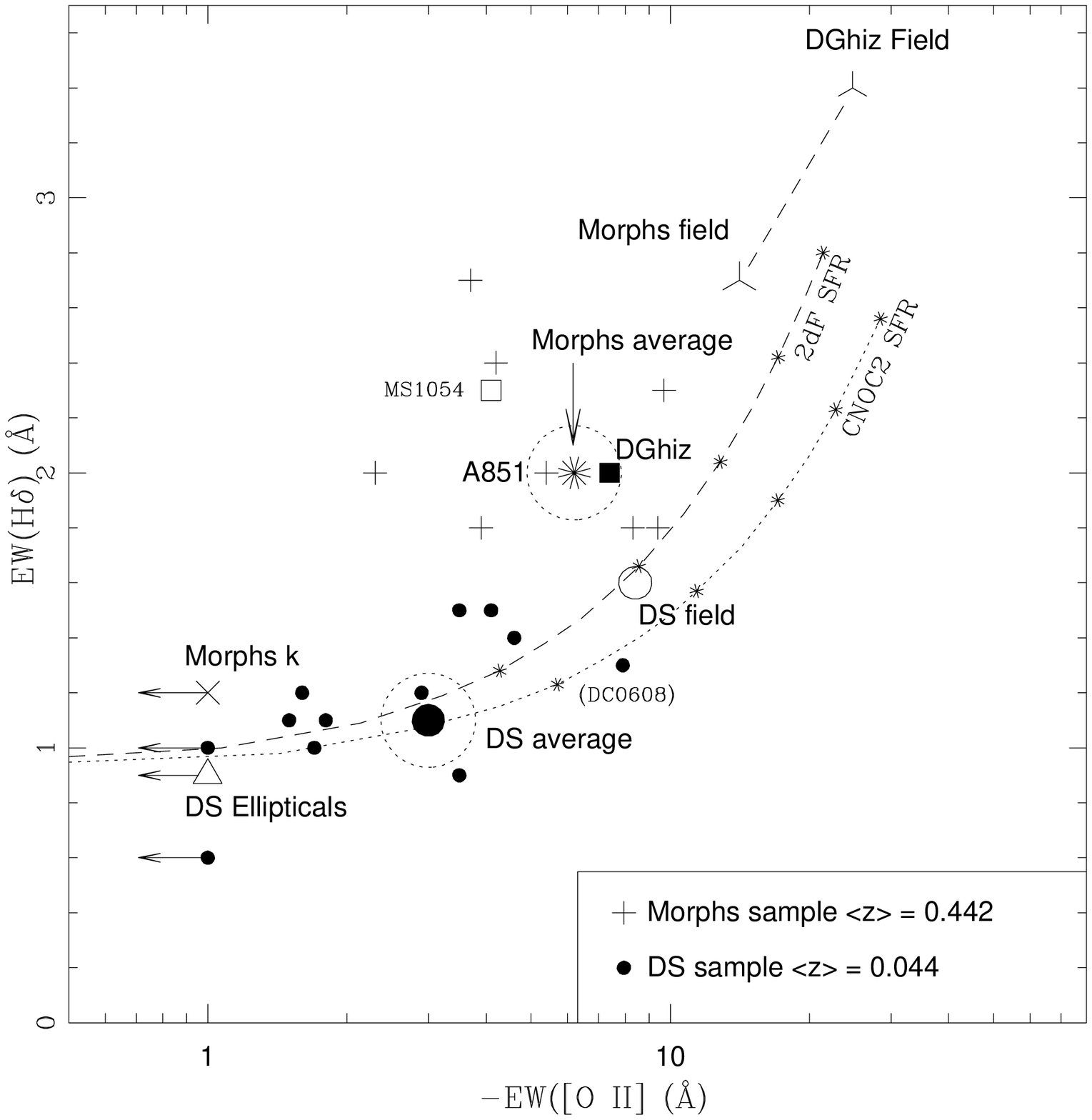}}
\caption{From Dressler et al. (2004). 
{\sl Left.} Composite spectra of five clusters at $z\sim 0.5$ (top five)
and five clusters at $z\sim 0$ (bottom five). The [O{\sc ii}] line is generally
more prominent in the high-z spectra. {\sl Right.} 
Equivalent widths of [O{\sc ii}]
versus $\rm H\delta$ as measured from composite spectra of clusters
at $z\sim 0.4-0.5$ (crosses) and clusters at $z\sim 0$ (filled dots).}
\end{figure}

The importance of the [O{\sc ii}] emission can also be assessed
from cluster composite
spectra, that are obtained summing up the light from all galaxies in a given
cluster to produce a sort of ``cluster integrated spectrum''
(Fig.~1, Dressler et al. 2004). 
As expected, the strength of [O{\sc ii}] in these composite spectra
displays a large cluster--to--cluster variation at any redshift, but
there is a tendency for the $z=0.5$ clusters to have on average a
stronger composite EW([O{\sc ii}]) than the clusters at $z=0$.

If numerous observations indicate that emission--line galaxies were
more prominent in clusters in the past than today, and if these
results are unsurprising given the evolution with $z$ of the star
formation activity in the general ``field'', {\it quantifying} this evolution
in clusters has proved to be very hard.  The fact that the
emission--line incidence varies strongly from a cluster to another at
all redshifts, and the relatively small samples of clusters studied in
detail at different redshifts, have so far hindered our progress in
measuring how the fraction of emission--line galaxies evolves with
redshift as a function of the cluster properties.

\begin{figure}[ht]
\centerline{\includegraphics[width=8cm]{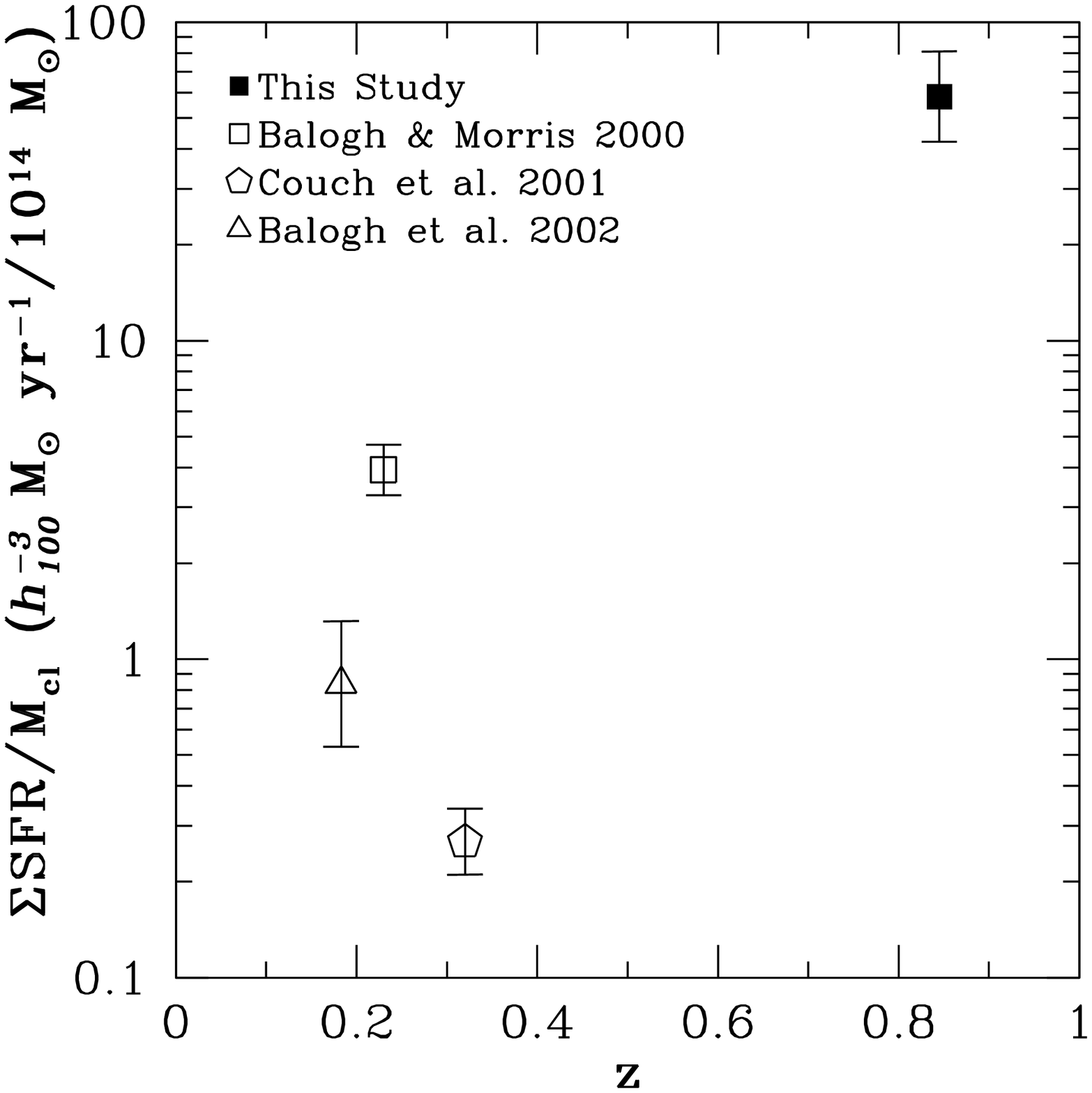}\hfill\includegraphics[width=8cm]{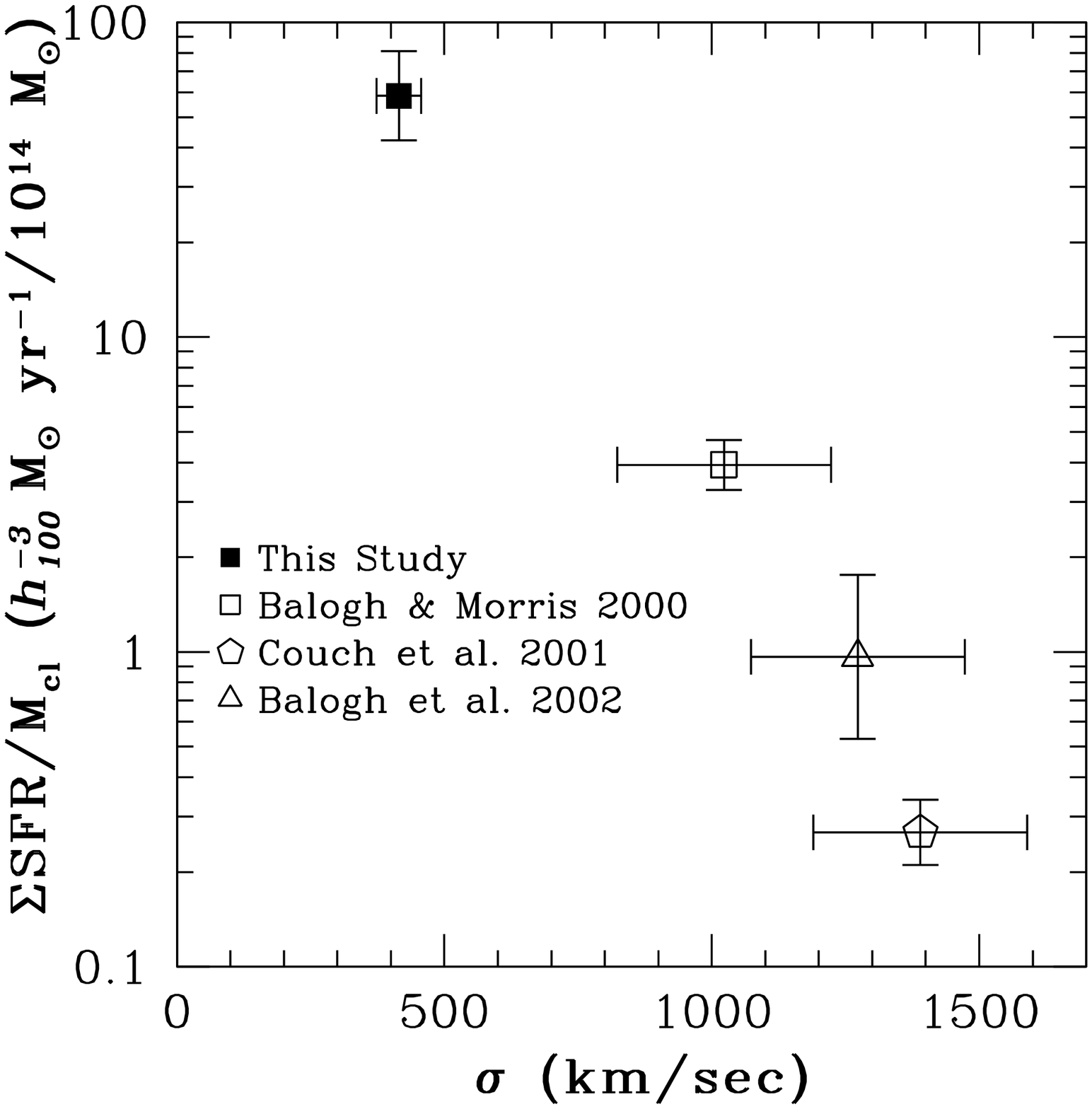}}
\caption{From Finn et al. (2004). Star formation rate per unit of cluster mass,
as measured from $\rm H\alpha$ narrow-band imaging, as a function of redshift
(left) and cluster velocity dispersion (right).}
\end{figure}

Discriminating between cosmic evolution and cluster--to--cluster
variance is a problem also for $\rm H\alpha$ cluster--wide
studies. Using narrow--band imaging or multiplex multislit
capabilities, a handful of clusters have been studied to date at $z
\ge 0.2$ (Couch et al. 2001, Balogh et al. 2002a, 
Finn et al. 2004, and submitted,
Kodama et al. 2004, Umeda et al. 2004).  These
studies have confirmed that the fraction of emission--line ($\rm
H\alpha$--detected, in this case) galaxies is lower in clusters than
in the field at similar redshifts, and have shown that the bright end
of the $\rm H\alpha$ luminosity function does not seem to depend
strongly on environment. As shown in Fig.~2, the number of clusters
studied is still insufficient to pin down the star formation rate
(SFR) per unit of cluster mass as a function of redshift AND of global
cluster properties such as the cluster velocity dispersion.

A word of caution is compulsory when using emission--lines and
assuming they provide an unbiased view of the evolution of the star
formation activity in cluster galaxies.  There are several indications
that dust extinction is in fact important and strongly distorts our
view of the star formation activity in at least some cluster galaxies.
Evidence for dust arises from optical spectroscopy itself, which finds
many dusty starbursting or star--forming galaxies with relatively weak
emission--lines both in distant clusters and in the field at similar
redshifts (Poggianti et al. 1999, Shioya et al. 2000, Poggianti
et al. 2001, Bekki et al. 2001). 
The radio--continuum detection of galaxies with
no optical emission lines (Smail et al. 1999, Miller \& Owen 2002) 
and mid-IR estimates of the star formation rate (Duc et
al. 2002, Coia et al. 2004 submitted, Biviano et al. 2004) indicate that even
the majority or all of the star formation activity of some cluster
galaxies can be obscured at optical wavelengths.  Whether taking into
account dust obscuration changes significantly the evolutionary
picture inferred from emission lines is still a critical open
question.

Finally, precious informations about emission--line galaxies can be
obtained from absorption--line spectra. In distant clusters, the
presence of galaxies with strong Balmer lines in absorption in their
spectra, and no emission lines, testifies that these are
post-starburst/post-starforming galaxies observed soon after their
star formation activity was interrupted and observed within 1-1.5 Gyr
from the halting (Dressler \& Gunn 1983, Couch \& Sharples 1987, see
Poggianti 2004 for a review). These galaxies have been found to
be proportionally more numerous in distant clusters than in the field at
comparable redshifts (Dressler et al. 1999, Poggianti et al. 1999,
Tran et al. 2003, 2004 ). Their spectral characteristics and their
different frequency as a function of the environment are a strong
indication for a truncation of the star formation activity related to
the dense environment.

\subsection{Galaxy morphologies}

In this section I summarize the results concerning the morphologies
of emission-line galaxies in distant clusters
obtained with the Wide Field and Planetary Camera 2. At the
time of writing, high--quality data of distant clusters have been
obtained with the Advanced Camera for Surveys by several groups, and
results should appear soon (Desai et al. in prep., Postman et al. in
prep.).

The HST images have revealed the presence of large numbers of spiral
galaxies in all distant clusters observed.  Comparing HST
morphologies and spectroscopy, it has been shown that
emission--line galaxies in distant clusters are for the great majority
{\it spirals} (Dressler et al. 1999).  The viceversa is not always
true: several of the cluster spirals, in fact, do not display any
emission line in their spectra, and both their spectra and their
colors indicate a lack of current star formation activity (Poggianti
et al. 1999, Couch et al. 2001, Goto et al. 2003).

These ``passive spirals'' might be an intermediate stage when
star--forming spirals are being transformed into passive S0
galaxies. A strong reason to believe that a significant fraction of
the spirals in distant clusters evolve into S0s comes from the
evolution of the Morphology-Density (MD) relation. The MD relation is
the observed correlation between the frequency of the various Hubble
types and the local galaxy density, normally defined as the projected
number density of galaxies within an area including its closest
neighbours. In clusters in the local Universe, the existence of this
relation has been known for a long time: ellipticals are frequent in
high density regions, while the fraction of spirals is high in low
density regions (Oemler 1974, Dressler et al. 1980).  At $z=0.4-0.5$,
an MD relation is already present, at least in concentrated clusters,
but it is {\it quantitatively} different from the relation at $z=0$:
the fraction of S0 galaxies at $z=0.5$ is much lower, at all
densities, than in clusters at $z=0$ (Dressler et al. 1997). The
fraction of S0s in clusters appears to increase towards lower
redshifts, while the proportion of spirals correspondingly decreases
(Dressler et al. 1997, Fasano et al. 2000).  Interestingly,
ellipticals are already as abundant at $z=0.5$ as at $z=0$. Adopting a
more conservative distinction between ``early-type'' (Es+S0s) and
late-type (spirals) galaxies, a similar evolution is found, with the
early-type fraction decreasing at higher redshifts (van Dokkum et
al. 2000, Lubin et al. 2002). First results at $z\sim 0.7-1.3$ seem to
indicate that between $z=0.5$ and $z=1$ what changes in the MD relation is 
only the occurrence of early-type galaxies in the very highest density regions
(Smith et al. 2004).

Alltogether, the findings described in this and the previous section
suggest that many galaxies have stopped forming stars in clusters
quite recently, as a consequence of environmental conditions 
switching off their star formation activity, and that many galaxies
have morphologically evolved from late to early type galaxies.
What can be the cause/causes for these changes?

\section{Physical processes}

The physical mechanisms that are usually considered when trying to
assess the influence of the environment on galaxy evolution can be
grouped in four main families:

\begin{enumerate}

\item
Mergers and strong galaxy-galaxy interactions (Toomre \& Toomre 1972,
Hernquist \& Barnes 1991, see Mihos 2004 for
a review).  These are most efficient when the relative velocities
between the galaxies are low, thus are expected to be especially
efficient in galaxy groups.

\item
Tidal forces due to the cumulative effect of many weaker encounters
(also known as ``harassment'') (Richstone 1976, Moore et al. 1998).
These are expected to be especially important in clusters, and particularly
on smaller / lower mass galaxies.

\item
Gas stripping - Interactions between the galaxy and the inter-galactic
medium (IGM) (Gunn \& Gott 1972, Quilis et al. 2000).  
The interstellar medium of a galaxy can be stripped via
various mechanisms, including viscous stripping, thermal evaporation
and -- the most famous member of this family -- ram pressure
stripping. Ram pressure can be efficient when the IGM gas density is
high and the relative velocity between the galaxy and the IGM is high,
and such conditions are expected to be met especially in the very
central regions of cluster cores.

\item
Strangulation (also known as starvation, or suffocation) (Larson,
Tinsley \& Caldwell 1980, Bower \& Balogh 2004). Assuming galaxies possess an
envelope of hot gas that can cool and feed the disk with fuel for star
formation, the removal of such reservoir of gas is destined to inhibit
further activity once the disk gas is exhausted. 
In semi-analytic models, for example, the gas halo is assumed to
be removed when a galaxy enters as satellite in a more massive
dark matter halo.

\end{enumerate}

Note that while stripping gas from the disk induces a truncation of
the star formation activity on a short timescale ($\sim 10^7$ yrs),
strangulation is expected to affect a galaxy star formation history
on a long timescale ($> 1$ Gyr) provoking a slowly declining activity
which consumes the disk gas
after the supply of cooling gas has been removed.

The former two of these families of processes affect the galaxy
structure, thus morphology, in a direct way: the merger of two spirals
can produce an elliptical galaxy, and repeated tidal encounters can
change a late--type into an early-type galaxy. The latter two
families, instead, act on the gas content of galaxies, hence their
star formation activity, and can modify their morphologies in an
indirect way: once star formation is halted in a disk,
this can fade significantly, the bulge-to-disk relative importance can
change and the galaxy appearance and morphology can appear
significantly modified.

\section{Low redshift}

Numerous excellent works have been carried out on a single cluster or samples
of clusters in the local Universe.
Summarizing them is beyond the scope of this paper, and the reader can
find a comprehensive review in Gavazzi \& Boselli  (in prep.). 
I have chosen to mention two low-redshift results here, to compare and
contrast them with the results at higher redshifts: the observed trends 
of star formation with local environment, and the gas/SF distribution
within galaxies.


\subsection{Trends of star formation with local environment}

It has been known for a long time that in the nearby Universe also the
average star formation activity correlates with the local density: in
higher density regions, the mean star formation rate per galaxy is
lower.  This is not surprising, given the existence of the MD
relation: the highest density regions have proportionally more
early-type galaxies devoid of current star formation.

Interestingly, the correlation between mean SF and local density
extends to very low local densities, comparable to those found at the
virial radius of clusters, and such a correlation exists also outside
of clusters (Lewis et al. 2002, Gomez et al. 2003). Again, this seems
to parallel the fact that an MD relation is probably existing in all
environments, and it has been observed in clusters of all types
(Dressler et al. 1980) and groups (Postman \& Geller 1984) -- though
the MD relation is {\it not the same} in all environments, e.g. in
concentrated vs. irregular, high- vs. low- $L_X$ clusters (Dressler et
al. 1980, Balogh et al. 2002b).

A variation in the mean SF/galaxy with density can be due either to a
difference in the {\it fraction of star-forming galaxies}, or in the
{\it star formation rates} of the star-forming galaxies, or a
combination of both.  In a recent paper, Balogh et al. (2004a) have
shown that the distribution of $\rm H\alpha$ equivalent widths (EW) in
star-forming galaxies does not depend strongly on the local density,
while the fraction of star-forming galaxies is a steep function of the
local density, in all environments.  Again, a dependence on
the {\it global environment} is observed, in the sense that, at a given 
local density, the fraction of emission-line galaxies is slightly lower
in environments with high density on large scales ($\sim 5$ Mpc) (but
see Kauffmann et al. 2004 for an opposite result).


The fact that a relation between star formation and density is
observed also outside of clusters has often been interpreted as a sign
that the environment starts affecting the star formation activity of
galaxies (provoking a decline in star formation in galaxies that
if isolated would continue forming stars) at relatively low densities,
when a galaxy becomes part of a group.  Personally, I believe the
existence of such a correlation is more probably the result of 
a correlation between initial conditions (galaxy mass and/or local environment
very early on, at the time the first stars formed in galaxies)
and type of galaxy formed. The exact shape of the correlation, instead,
is probably influenced by transformations happening in galaxies
when they enter a different environment.

\subsection{Gas content and gas/SF distribution within galaxies}

In order to understand what happens to galaxies in clusters, two
crucial pieces of information are 1) the gas content of cluster
galaxies and 2) the spatial distribution of the gas and of the star
formation activity within each galaxy.

It has been several years since it became evident that many spirals in
clusters are deficient in HI gas compared to similar galaxies in the
field (Giovanelli \& Haynes 1985, Cayatte et al. 1990, 
see van Gorkom 2004 for a review).  Most (but not all)
of the HI deficient spirals are found at small distances from the
cluster centre. In the central regions of clusters, the sizes of the
HI disks are smaller than the optical disks, and a spatial
displacement between the HI and the optical occurs in several cases
(Bravo-Alfaro et al. 2000).  
The fraction of HI--deficient spirals increases going towards
the cluster centre, and a correlation is observed between deficiency
and orbital parameters: more deficient galaxies tend to be on radial
orbits (Solanes et al. 2001).

All of these findings strongly suggest that ram pressure stripping, or
at least gas stripping in general, plays an important role (see also
Bravo-Alfaro and Solanes contributions in these proceedings).  On the
other hand, the work from Solanes et al. (2001) has unexpectedly shown
that the HI deficiency is observed out to 2 Abell radii.  This result
has raised the question whether the origin of the HI deficiency in the
cluster outskirts can be consistent with the ram pressure scenario and
whether can be simply due to effects such as large distance errors or
rebounding at large clustercentric distances of galaxies that have
gone through the cluster center (Balogh, Navarro \& Morris 2000, Mamon
et al. 2004, Moore et al. 2004, Sanchis et al. 2004).

\begin{figure}[ht]
\includegraphics[width=12cm]{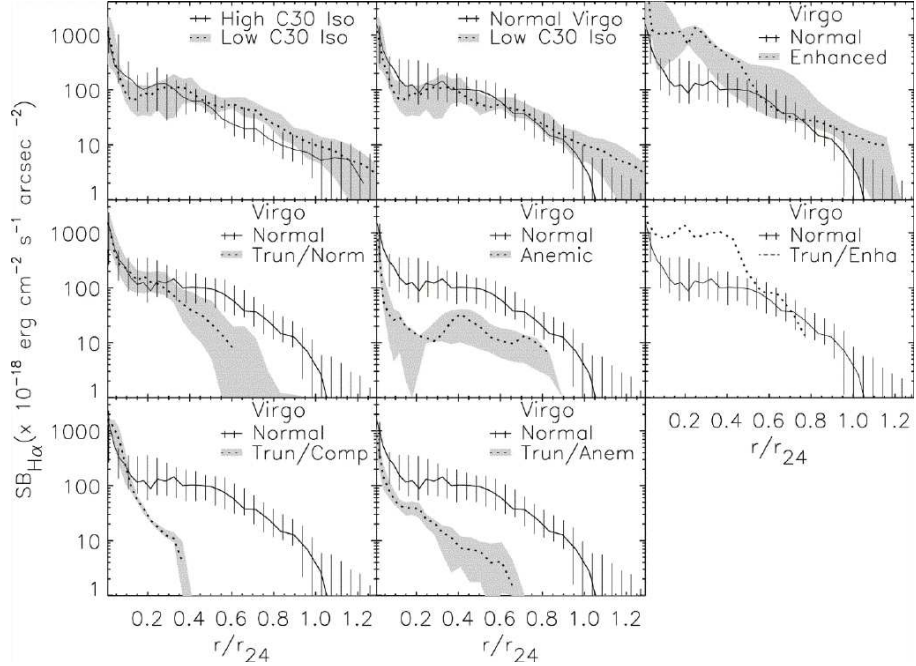}
\caption{From Koopmann \& Kenney 2004. Median $\rm H\alpha$ radial profiles 
for galaxies grouped in different classes according to their $\rm H\alpha$
distribution. The first two panels show profiles that can be considered
``normal'' (reproduced also in other panels). Truncated, anemic, enhanced 
classes are shown in all the other panels.}
\end{figure}

Recent works have yielded a census of the spatial
distribution of the star formation (as observed in $\rm H\alpha$
emission) within cluster galaxies. These works have shown that the
majority of the cluster galaxies have peculiar $\rm H\alpha$
morphologies, compared to field galaxies
(Moss \& Whittle 1993, 2000, Koopmann
\& Kenney 2004, Vogt et al. 2004). More than half
of the spirals in the Virgo cluster, for example, have $\rm H\alpha$
radial profiles truncated from a certain radius on, while others have
$\rm H\alpha$ suppressed throughout the disk, and in some cases,
enhanced (Fig.~3) (Koopmann \& Kenney 2004). On a sample of 18 nearby
clusters, similar classes of objects are observed: spirals with
truncated $\rm H\alpha$ emission and HI gas on the leading edge of the
disk; spirals stripped of their HI with their star formation confined
to the inner regions; and quenched spirals, in which the star
formation is suppressed throughout the disk (Vogt et al. 2004).
From these works, gas stripping appears a very important factor in
determining both the gas content and the star formation activity of
cluster spirals, though tidal effects are also found to be significant
(Moss \& Whittle 2000, Koopmann \& Kenney 2004).
Spirals in clusters thus appear to be heavily affected by the
environment, and to be observed in different stages of their likely
transformation from infalling star-forming spirals to cluster S0s.

It is apparently hard to reconcile the peculiar $\rm H\alpha$ morphologies of
spirals in nearby clusters with the fact that the distribution of $\rm
H\alpha$ equivalent widths (EWs) for blue galaxies in the 2dF and Sloan
does not seem to
vary significantly in clusters, groups and field (Fig.~9 in Balogh et
al. 2004). This apparent contradiction still awaits an
explanation.

\section{Trends with galaxy mass and downsizing effect}

How do the results described above depend on the galaxy mass? Does
the environmental dependence of galaxy properties change with galaxy mass?

It has always been known that fainter, lower mass galaxies on average are bluer
than higher mass galaxies, and on average have 
more active star formation. The references
to old and new papers showing this could easily fill this review.

\begin{figure}[ht]
\includegraphics[width=13cm]{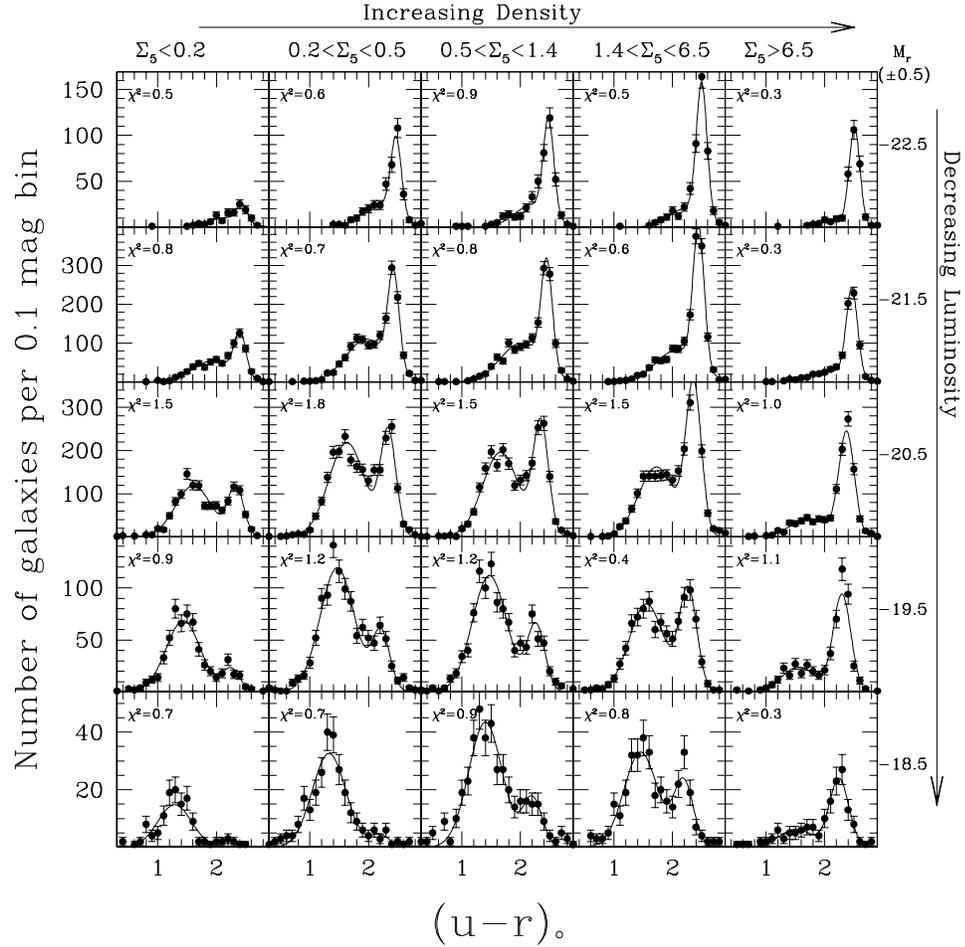}
\caption{From Balogh et al. 2004b. Galaxy color distribution from Sloan
as a function of luminosity and local density.}
\end{figure}

Recently, the interest in this well established observational result
has grown, and its implications have been more and more appreciated.
In {\it all} environments, lower mass galaxies have on average a more
protracted star formation history. This implies that, on average,
going to lower redshifts, the maximum luminosity/mass of galaxies with
significant star formation activity progressively decreases.  This
``downsizing effect'', observed in and outside of clusters, indicates
an ``anti-hierachical'' history for the star formation in galaxies,
which parallels a similar effect observed for AGNs (Cristiani et
al. 2004, Shankar et al. 2004).  The downsizing effect is thus another
effect {\it besides} any environmental effect (see e.g. Fig.~4, 
and Kauffmann et al. 2003, 2004), and the dependence on
the galaxy mass/luminosity cannot be ignored when trying to trace
evolutionary effects.

In clusters, innumerable results have shown the existence of a
downsizing effect (Smail et al. 1998, 
Gavazzi et al. 2002, De Propris et al. 2003, Tran et al. 2003,
De Lucia et al. 2004, Kodama et al. 2004, Poggianti et al. 2004, to name a 
few). A direct observation of this effect at high redshift is shown
in Fig.~5  and illustrates the consequence
of downsizing on the characteristics of the color-magnitude red sequence
in clusters. A deficiency of faint red galaxies is observed compared
to Coma in all four clusters studied, despite of the variety of cluster
properties. The red luminous galaxies are already in place
on the red sequence at $z\sim 0.8$, while a significant fraction of
the faint galaxies must have stopped forming stars and, consequently, moved
on to the red sequence at lower redshifts.

\begin{figure}[ht]
\centerline{\includegraphics[width=5cm]{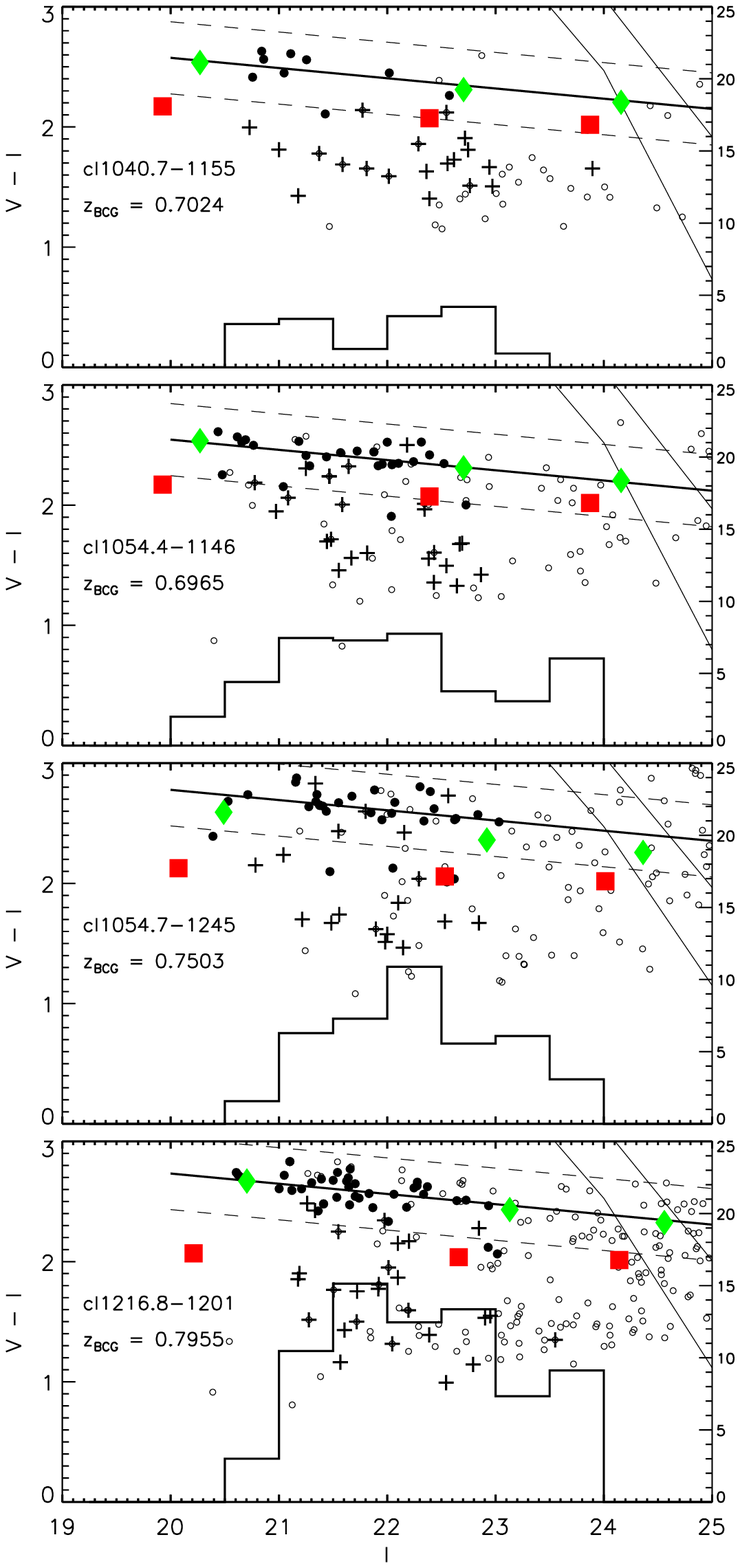}\hfill\includegraphics[width=8cm]{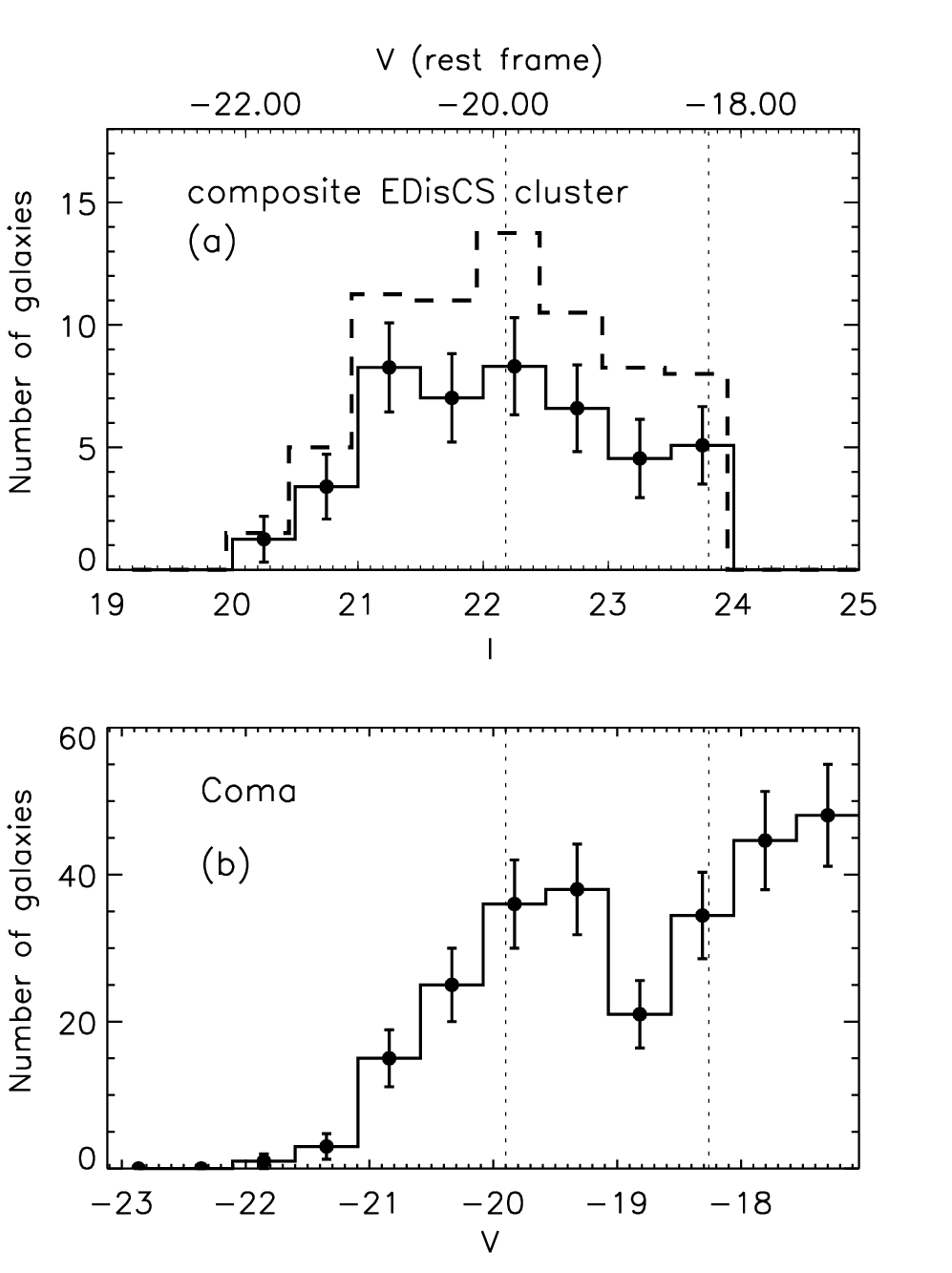}}
\caption{From De Lucia et al. 2004. {\sl Left.} 
Color-magnitude diagrams of four
clusters at $z=0.7-0.8$ from the ESO Distant Cluster Survey. Histograms
represent the magnitude distribution of galaxies within $3\sigma$ from 
the red sequence. {\sl Right.} 
The magnitude distribution of red sequence
galaxies at $z\sim 0.75$ (top) is compared with the one of red galaxies
in the Coma cluster (bottom). High-z clusters exhibit a clear deficit of 
low-luminosity passive red galaxies compared to Coma and other nearby 
clusters.}
\end{figure}

\section{Conclusions and speculations}

A number of important issues have not been considered here due to
page limits, but should be included in any complete review
of emission-line galaxies in clusters, in particular:
%
%
a description of the IR and radio-continuum
studies, the Tully-Fisher relation of cluster versus field spirals,
the spatial distribution and kinematics of various types of
galaxies, the link between the star-formation activity and cluster
substructure, the luminosity function of emission-- and
non--emission line galaxies as a function of the environment, and the
line emission from AGNs.

By now, mostly thanks to the scientific debate of the latest years, it
is widespread wisdom that we are investigating a three-dimensional
space, whose axis are redshift, environment and galaxy mass.  In fact,
the evolutionary histories and, in particular, the star formation
activity of galaxies have similar (increasing) trends as a function
of (higher) redshift, (lower) galaxy mass and (lower) density/mass of
the environment. While observationally we are beginning to fill this
3-parameter space, the great theoretical challenge is to help
comprehend from a physical point of view why this is the history of
our Universe.

The results mentioned above do not yet match together into one, coherent
picture, and in some cases they might be apparently clashing with each
other. The observational results in distant clusters, both those regarding
the star formation activity and the morphologies, point to a strong 
evolution. Many galaxies have stopped forming stars in clusters
quite recently, and the morphology of many disk galaxies must
have changed. Dense environments seem to accelerate the
transformation of star-forming late-type galaxies into passive
early-type galaxies. The local large redshift surveys have highlighted
the existence of trends of galaxy properties with local density
that exist in all environments. This, as well as several 
other lines of evidence, suggests that whatever induces the existence
of a morphology-density and of a star-formation-density relation
is not effective {\it solely} in clusters. On the other hand,
there is now solid evidence (and the gas/star formation results
in nearby clusters are most convincing in this respect) that gas stripping
is at work and affects {\it most} of the emission-line galaxies in clusters,
(or, at least, most of those galaxies in clusters that today still have
emission lines).

Twenty years ago, an expression often used in this field was
``nature or nurture?''. Nowadays, especially within the context 
of a hierarchically growing Universe, the distinction between nature
and nurture has become very subtle, also from a philosophical point of view. 

The example of the most massive ellipticals is instructive in this respect.
They are found in the cluster cores, and the current cosmological paradigm
tells us that they were in the highest density peaks of the primordial 
fluctuations, thus have always been in the highest density regions,
since very high redshifts.
Therefore, at least ``part'' of the morphology-density relation (and of
the star-formation density (SFD) relation, since their stars all formed
very early on) must have been in place at $z>3$.
Is it more correct to call this a ``primordial effect'' -- because
established at very high redshift -- or ``environmental'' -- because
related to the local density of the environment? 
but the environment of these ellipticals at $z=10$ is strongly related
to their environment today: can we speak of ``galaxy destiny''??

On the other hand, we know that the morphology-density relation in clusters
has evolved with redshift, at least as far as the disk galaxies
are concerned, and we can directly observe its evolution.
Therefore, we know that while the existence of {\it a} morphology-density
relation can be traced back to the first epoch of galaxy formation,
its evolution, and thus its exact shape at any time, depends
on the subsequent evolution, likely strongly influenced by the environment.
Is the {\it existence} of the MD (and SFD) relation primordial, and its 
{\it evolution} environmental?

Perhaps also we, as behavioural psychologists (Ridley, 2003),
are coming to realize that nature and nurture are so intertwined
that they become, 
to a certain extent,
indistiguishable and unseparable. 




\begin{acknowledgments}
I wish to thank Mariano Moles for the kindly provided financial support
and the organizers of this JENAM meeting and, in particular, of the
Session on ``The life of galaxies'' for their invitation and support,
which allowed me to participate to a very pleasant and interesting
meeting.

\end{acknowledgments}

\end{document}